\documentclass[conference]{IEEEtran}
\IEEEoverridecommandlockouts
\usepackage{amsmath,amssymb,amsfonts}

\usepackage{graphicx}
\usepackage{booktabs}
\usepackage{xcolor}
\usepackage[version=4]{mhchem}
\usepackage{hyperref}
\usepackage{microtype}
\usepackage[style=numeric, giveninits=true, hyperref, doi=true, isbn=true, maxbibnames=19, date=year, sorting=none, backend=biber]{biblatex}
\usepackage{siunitx}
\renewbibmacro{in:}{}

\renewbibmacro*{issue+date}{}% suppress "(year)" after volume(number)
\renewbibmacro*{note+pages}{%
  \printfield{note}%
  \setunit{\bibpagespunct}%
  \printfield{pages}%
  \setunit{\addcomma\space}%
  \usebibmacro{date}%
  \newunit}

\DeclareSourcemap{
  \maps[datatype=bibtex]{
    \map{
 \pertype{article}
 \step[fieldset=issn, null]
    }
  }
}
\DeclareFieldFormat*{title}{#1\addspace}
\DefineBibliographyStrings{english}{%
  page  = {},
  pages = {},
}
\renewbibmacro*{volume+number+eid}{%
  \printfield{volume}%
  \printfield{number}%
  \setunit{\addcomma\space}%
  \printfield{eid}}
\DeclareFieldFormat[article]{volume}{\mkbibbold{#1}}
\DeclareFieldFormat[article]{number}{\mkbibparens{#1}}

\DeclareFieldFormat[inproceedings]{volume}{\mkbibbold{#1}}
\DeclareFieldFormat[inproceedings]{number}{\mkbibparens{#1}}

\DeclareFieldFormat[inproceedings]{booktitle}{\mkbibemph{#1}\nopunct}
\def\BibTeX{{\rm B\kern-.05em{\sc i\kern-.025em b}\kern-.08em T\kern-.1667em\lower.7ex\hbox{E}\kern-.125emX}}

\usepackage{tikz}
\usepackage{pgfplots}
\usepackage{pgfplotstable}
\usepackage[official]{eurosym}
\usetikzlibrary{backgrounds,patterns}
\pgfplotsset{compat=1.18}
\usepgfplotslibrary{groupplots}

\makeatletter
\newcommand{\resetstackedplots}{\pgfplots@stacked@isfirstplottrue}
\makeatother

\pgfplotsset{
  only dk/.style={restrict expr to domain={\coordindex}{0:5}},
  only fi/.style={restrict expr to domain={\coordindex}{6:11}},
  only no/.style={restrict expr to domain={\coordindex}{12:17}},
  only se/.style={restrict expr to domain={\coordindex}{18:23}},
}

\definecolor{cNuclear}{RGB}{255,0,0}
\definecolor{cHydro}{RGB}{35,140,130}
\definecolor{cCHP}{RGB}{140,90,80}
\definecolor{cCCGT}{RGB}{170,85,20}
\definecolor{cOnwind}{RGB}{30,90,200}
\definecolor{cOffwind}{RGB}{110,160,230}
\definecolor{cSolarHSAT}{RGB}{255,175,0}
\definecolor{cSolarFixed}{RGB}{255,215,0}
\definecolor{cSolarRooftop}{RGB}{255,235,140}

\definecolor{cDist}{RGB}{70,130,40}
\definecolor{cIndustry}{RGB}{40,40,90}
\definecolor{cHeatPump}{RGB}{0,170,0}
\definecolor{cBEV}{RGB}{180,230,60}
\definecolor{cHydrogen}{RGB}{255,0,255}
\definecolor{cMeth}{RGB}{0,220,220}
\definecolor{cAmmo}{RGB}{0,120,120}
\definecolor{cExport}{RGB}{70,70,70}

\definecolor{cThermal}{RGB}{120,120,120}
\definecolor{cSolar}{RGB}{240,180,0}

\definecolor{cDataCentre}{RGB}{140,174,140}
\definecolor{cTransport}{RGB}{182,200,209}
\definecolor{cIndustry}{RGB}{133,107,117}
\definecolor{cRealised}{RGB}{43,128,149}

\pgfplotstableread[col sep=comma]{
Year,Households,Industry,Transport,DataCentre,Hydrogen
2025,18,20,2,0,0
2030,20,23,7,17,14
2035,0,0,0,0,0
2040,24,26,14,29,45
2045,0,0,0,0,0
2050,26,29,20,28,130
}\consDKngdp

\pgfplotstableread[col sep=comma]{
Year,Households,Industry,Transport,DataCentre,Hydrogen
2025,47,40,2,0,0
2030,54,69,0,0,0
2035,0,0,0,0,0
2040,55,84,7,0,65
2045,0,0,0,0,0
2050,58,96,11,0,90
}\consFIngdp

\pgfplotstableread[col sep=comma]{
Year,Households,Industry,Transport,DataCentre,Hydrogen
2025,78,56,5,0,0
2030,76,70,10,6,0
2035,0,0,0,0,0
2040,73,76,23,14,9
2045,0,0,0,0,0
2050,72,78,27,18,26
}\consNOngdp

\pgfplotstableread[col sep=comma]{
Year,Households,Industry,Transport,DataCentre,Hydrogen
2025,84,47,3,3,0
2030,87,67,9,3,10
2035,0,0,0,0,0
2040,90,77,20,5,44
2045,0,0,0,0,0
2050,92,78,26,9,60
}\consSEngdp

\pgfplotstableread[col sep=comma, columns/Country/.style={string type}
]{balances_nordics_by_country_pgfplots.csv}\balancesAll

\pgfplotstableread[col sep=comma]{
Year,Hydro,Nuclear,Thermal,WindOnshore,WindOffshore,Solar
2025,0.0,0.0,6.5,5.7,2.9,5.7
2030,0.0,0.0,4.1,6.5,4.1,21.0
2035,0,0,0,0,0,0
2040,0.0,0.0,2.4,5.9,24.2,37.3
2045,0,0,0,0,0,0
2050,0.0,0.0,2.4,5.3,44.4,44.7
}\capDKngdp

\pgfplotstableread[col sep=comma]{
Year,Hydro,Nuclear,Thermal,WindOnshore,WindOffshore,Solar
2025,3.6,3.6,9.3,10.1,0.7,1.4
2030,3.3,2.7,8.3,18.6,0.0,9.5
2035,0,0,0,0,0,0
2040,3.3,2.7,8.3,46.1,0.0,20.1
2045,0,0,0,0,0,0
2050,3.3,2.7,8.3,50.0,0.0,28.1
}\capFIngdp

\pgfplotstableread[col sep=comma]{
Year,Hydro,Nuclear,Thermal,WindOnshore,WindOffshore,Solar
2025,39.5,0.0,0.0,6.5,0.0,2.2
2030,34.0,0.0,0.0,5.0,1.2,5.6
2035,0,0,0,0,0,0
2040,39.0,0.0,0.0,6.5,3.8,13.3
2045,0,0,0,0,0,0
2050,39.3,0.0,0.0,8.0,5.9,21.9
}\capNOngdp

\pgfplotstableread[col sep=comma]{
Year,Hydro,Nuclear,Thermal,WindOnshore,WindOffshore,Solar
2025,19.4,8.6,5.7,19.4,0.0,7.9
2030,16.3,7.1,4.4,20.4,0.0,4.7
2035,0,0,0,0,0,0
2040,16.3,5.0,4.4,25.4,1.5,10.9
2045,0,0,0,0,0,0
2050,16.3,8.0,4.1,28.1,1.8,17.2
}\capSEngdp

\pgfplotstableread[col sep=comma, columns/Country/.style={string type}
]{installed_capacities_nordic_by_country_pgfplots.csv}\capsAll

\begin{document}

\title{Modelling the Nordic power system expansion with PyPSA-Eur: a comparison with the Nordic Grid Development Perspective 2025
\thanks{\revise{Financed by the Nordic Network of Living Labs (NNLL) project, grant number NordForsk 242039.} Supported by the FME SOLAR project, grant number NRC 350244}
}

\author{
\IEEEauthorblockN{1\textsuperscript{st} Leonardo Rydin Gorjão}
\IEEEauthorblockA{\textit{Environmental Sciences}\\
\textit{Open University of the Netherlands} \\
Heerlen, The Netherlands \&\\
\textit{Norwegian University of Life Sciences (NMBU)}\\
Ås, Norway \\
leonardo.rydingorjao@ou.nl}
\and
\IEEEauthorblockN{2\textsuperscript{nd} Heidi S. Nygård}
\IEEEauthorblockA{\textit{Norwegian University of Life Sciences (NMBU)}\\
Ås, Norway \\
}}

% \author{
% \IEEEauthorblockN{1\textsuperscript{st} Anonymous Author}
% \and
% \IEEEauthorblockN{2\textsuperscript{nd} Anonymous Author}
% }

\newcommand{\revise}[1]{\textcolor{black}{#1}}

\maketitle

\begin{abstract}
We compare the Nordic Grid Development Perspective 2025 (NGDP) with a least-cost transition pathway for the Nordic power system obtained with PyPSA-Eur for the period 2025--2050 under a net-zero \ce{CO2} constraint.
The model co-optimises generation, storage, and transmission expansion with hourly system operation while constraining the Nordic generation mix to remain consistent with NGDP milestone ranges.
Overall, the optimised capacity pathways reproduce the broad NGDP expansion of wind and solar power, with Denmark emerging as the main locus of offshore wind and solar deployment, Finland as a major site for onshore wind expansion, and Norway and Sweden relying more strongly on existing hydropower and nuclear capacity.
The main discrepancy between the two pathways does not lie in the broad build-out of variable renewables, but in the timing and role of hydrogen.
Whereas the NGDP treats hydrogen as an early and steadily increasing component of industrial demand and system development already from 2030, PyPSA-Eur delays substantial hydrogen deployment until 2045--2050, when it becomes cost-effective primarily as a flexibility and seasonal storage option in a strongly \revise{variable-renewable-energy-}dominated system.
This contrast suggests that early hydrogen build-out in the Nordics is not a robust outcome of cost optimisation alone, but depends strongly on assumptions regarding industrial demand growth, sector coupling, and anticipatory infrastructure planning.
We therefore argue that hydrogen remains the principal strategic uncertainty in the Nordic transition, with the NGDP representing a more anticipatory planning perspective and PyPSA-Eur a more conservative least-cost timing of deployment.
\end{abstract}


\section{Introduction}\label{sec:introduction}

%~\cite{IEA2025,IRENA2024}

The drive to decarbonise the energy system and meet the net-zero target by 2050 create a strong push towards implementing renewable energy sources in Europe~\cite{IPCC2023,EuropeanCommission2019GreenDeal,EU2021ClimateLaw}.
The Nordics, although boasting the lowest carbon emissions in the European power system, face an expected significant increase in electricity demand, driven by electrification and new industrial loads~\cite{NGDP2025}.
No major expansion of either hydropower generation or nuclear power is foreseen in the near future, which leaves open a strong role for wind and solar, complemented by flexibility, strengthened transmission, and demand-side management measures.
Yet, the implementation of variable renewable energy (VRE) sources in the Nordics is challenged by spatially heterogeneous resources, seasonal variability, and a transmission grid that must accommodate both new generation areas and new demand centres~\cite{NGDP2025}.
The nuclear power base in Sweden and Finland can serve as a strong source of stability and baseload, and the fast-responding and vast hydropower resources in Norway can provide flexibility and balancing services, but matching generation and demand will be left to a combination of wind and solar generation, energy storage, a more developed demand response, and much more cross-border exchange~\cite{NGDP2025}.
Moreover, energy storage technologies, such as hydrogen electrolysis, are expected to play a significant role in providing seasonal storage and balancing services, but their deployment, integration, and impact into the system remain to be uncovered~\cite{Neumann2025,Fleiter2025}.

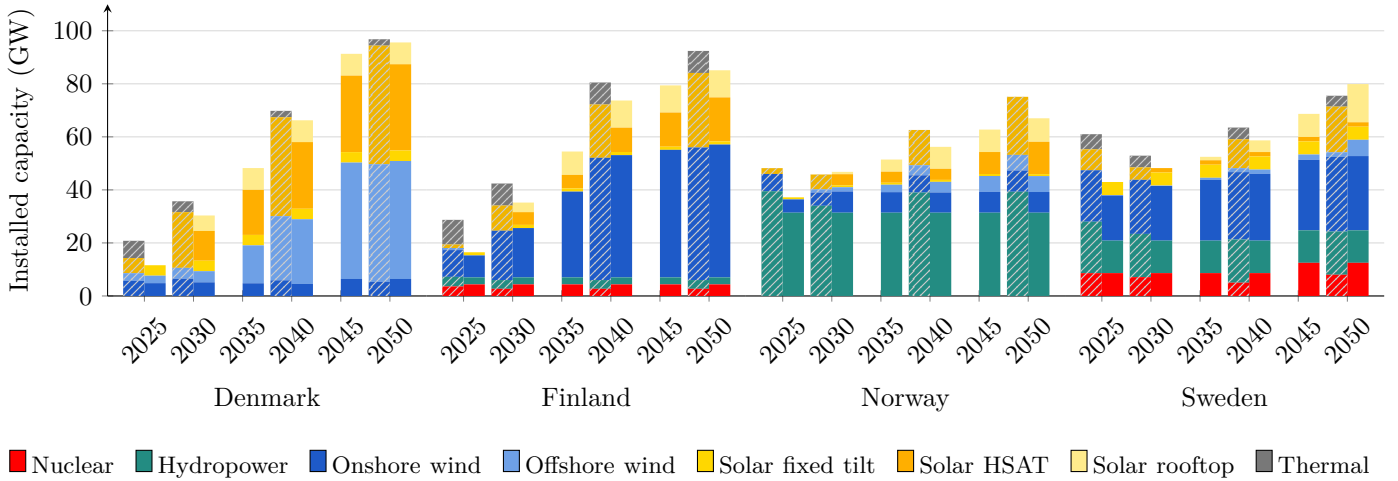
\begin{figure*}[ht]
\centering
\hspace*{-0.2cm}\begin{tikzpicture}

\pgfplotsset{
  sidebysidecap/.style={
    ybar stacked,
    bar width=8pt,
    symbolic x coords={2025,2030,2035,2040,2045,2050},
    xtick=data,
    x tick label style={rotate=45},
    x axis line style={-},
    ymin=0, ymax=110,
    ytick={0,20,40,60,80,100},
    ymajorgrids,
    major grid style={black!15},
    axis x line=bottom,
    axis y line=left,
    enlarge x limits=0.15,
  x axis line style={-},
  }
}

\begin{groupplot}[
  group style={
    group size=4 by 1,
    horizontal sep=0cm,
    ylabels at=edge left,
    yticklabels at=edge left,
  },
  width=0.32\linewidth,
  height=0.30\linewidth,
  sidebysidecap,
]

% =================================================================
% Denmark
% =================================================================
\nextgroupplot[
  xlabel={Denmark},
  ylabel={Installed capacity (\unit{GW})},
  legend to name=combinedcaplegend,
  legend columns=8,
  legend cell align=left,
  legend style={
 draw=none, fill=none, font=\small,
    /tikz/every even column/.append style={column sep=0.6em},
  },
]
% PyPSA bars (shifted right, solid)
\addplot[bar shift=4pt, fill=cNuclear, draw=none, only dk] table[x=Year, y={Nuclear}] {\capsAll}; \addlegendentry{Nuclear}
\addplot[bar shift=4pt, fill=cHydro, draw=none, only dk] table[x=Year, y={Hydroelectricity}] {\capsAll};  \addlegendentry{Hydropower}
\addplot[bar shift=4pt, fill=cOnwind, draw=none, only dk] table[x=Year, y={Onshore wind}] {\capsAll}; \addlegendentry{Onshore wind}
\addplot[bar shift=4pt, fill=cOffwind, draw=none, only dk] table[x=Year, y={Offshore wind}] {\capsAll};     \addlegendentry{Offshore wind}
\addplot[bar shift=4pt, fill=cSolarFixed,  draw=none, only dk] table[x=Year, y={Solar fixed tilt}] {\capsAll};  \addlegendentry{Solar fixed tilt}
\addplot[bar shift=4pt, fill=cSolarHSAT, draw=none, only dk] table[x=Year, y={Solar HSAT}] {\capsAll};  \addlegendentry{Solar HSAT}
\addplot[bar shift=4pt, fill=cSolarRooftop, draw=none, only dk] table[x=Year, y={Solar rooftop}] {\capsAll};     \addlegendentry{Solar rooftop}
\addlegendimage{fill=cThermal, draw=none}    \addlegendentry{Thermal}
\resetstackedplots
% NGDP bars (shifted left, hatched)
\addplot[bar shift=-4pt, fill=cNuclear, draw=none, postaction={pattern=north east lines, pattern color=black!20}] table[x=Year, y=Nuclear] {\capDKngdp}; %\addlegendentry{Nuclear}
\addplot[bar shift=-4pt, fill=cHydro,  draw=none, postaction={pattern=north east lines, pattern color=black!20}] table[x=Year, y=Hydro] {\capDKngdp};  %\addlegendentry{Hydropower}
\addplot[bar shift=-4pt, fill=cOnwind, draw=none, postaction={pattern=north east lines, pattern color=black!20}] table[x=Year, y=WindOnshore] {\capDKngdp};  %\addlegendentry{Onshore wind}
\addplot[bar shift=-4pt, fill=cOffwind, draw=none, postaction={pattern=north east lines, pattern color=black!20}] table[x=Year, y=WindOffshore] {\capDKngdp}; %\addlegendentry{Offshore wind}
\addplot[bar shift=-4pt, fill=cSolar,  draw=none, postaction={pattern=north east lines, pattern color=black!20}] table[x=Year, y=Solar] {\capDKngdp};  %\addlegendentry{Solar}
\addplot[bar shift=-4pt, fill=cThermal, draw=none, postaction={pattern=north east lines, pattern color=black!20}] table[x=Year, y=Thermal] {\capDKngdp}; %\addlegendentry{Thermal}

% =================================================================
% Finland
% =================================================================
\nextgroupplot[
  xlabel={Finland},
  axis y line=left, y axis line style={opacity=0}, ytick style={draw=none},
]
\addplot[bar shift=-4pt, fill=cNuclear, draw=none, forget plot, postaction={pattern=north east lines, pattern color=black!20}] table[x=Year, y=Nuclear] {\capFIngdp};
\addplot[bar shift=-4pt, fill=cHydro,  draw=none, forget plot, postaction={pattern=north east lines, pattern color=black!20}] table[x=Year, y=Hydro] {\capFIngdp};
\addplot[bar shift=-4pt, fill=cOnwind, draw=none, forget plot, postaction={pattern=north east lines, pattern color=black!20}] table[x=Year, y=WindOnshore] {\capFIngdp};
\addplot[bar shift=-4pt, fill=cOffwind, draw=none, forget plot, postaction={pattern=north east lines, pattern color=black!20}] table[x=Year, y=WindOffshore] {\capFIngdp};
\addplot[bar shift=-4pt, fill=cSolar,  draw=none, forget plot, postaction={pattern=north east lines, pattern color=black!20}] table[x=Year, y=Solar] {\capFIngdp};
\addplot[bar shift=-4pt, fill=cThermal, draw=none, forget plot, postaction={pattern=north east lines, pattern color=black!20}] table[x=Year, y=Thermal] {\capFIngdp};
\resetstackedplots
\addplot[bar shift=4pt, fill=cNuclear, draw=none, only fi, forget plot] table[x=Year, y={Nuclear}] {\capsAll};
\addplot[bar shift=4pt, fill=cHydro, draw=none, only fi, forget plot] table[x=Year, y={Hydroelectricity}] {\capsAll};
\addplot[bar shift=4pt, fill=cOnwind, draw=none, only fi, forget plot] table[x=Year, y={Onshore wind}] {\capsAll};
\addplot[bar shift=4pt, fill=cOffwind, draw=none, only fi, forget plot] table[x=Year, y={Offshore wind}] {\capsAll};
\addplot[bar shift=4pt, fill=cSolarFixed,  draw=none, only fi, forget plot] table[x=Year, y={Solar fixed tilt}] {\capsAll};
\addplot[bar shift=4pt, fill=cSolarHSAT, draw=none, only fi, forget plot] table[x=Year, y={Solar HSAT}] {\capsAll};
\addplot[bar shift=4pt, fill=cSolarRooftop, draw=none, only fi, forget plot] table[x=Year, y={Solar rooftop}] {\capsAll};

% =================================================================
% Norway
% =================================================================
\nextgroupplot[
  xlabel={Norway},
  axis y line=left, y axis line style={opacity=0}, ytick style={draw=none},
]
\addplot[bar shift=-4pt, fill=cNuclear, draw=none, forget plot, postaction={pattern=north east lines, pattern color=black!20}] table[x=Year, y=Nuclear] {\capNOngdp};
\addplot[bar shift=-4pt, fill=cHydro,  draw=none, forget plot, postaction={pattern=north east lines, pattern color=black!20}] table[x=Year, y=Hydro] {\capNOngdp};
\addplot[bar shift=-4pt, fill=cOnwind, draw=none, forget plot, postaction={pattern=north east lines, pattern color=black!20}] table[x=Year, y=WindOnshore] {\capNOngdp};
\addplot[bar shift=-4pt, fill=cOffwind, draw=none, forget plot, postaction={pattern=north east lines, pattern color=black!20}] table[x=Year, y=WindOffshore] {\capNOngdp};
\addplot[bar shift=-4pt, fill=cSolar,  draw=none, forget plot, postaction={pattern=north east lines, pattern color=black!20}] table[x=Year, y=Solar] {\capNOngdp};
\addplot[bar shift=-4pt, fill=cThermal, draw=none, forget plot, postaction={pattern=north east lines, pattern color=black!20}] table[x=Year, y=Thermal] {\capNOngdp};
\resetstackedplots
\addplot[bar shift=4pt, fill=cNuclear, draw=none, only no, forget plot] table[x=Year, y={Nuclear}] {\capsAll};
\addplot[bar shift=4pt, fill=cHydro, draw=none, only no, forget plot] table[x=Year, y={Hydroelectricity}] {\capsAll};
\addplot[bar shift=4pt, fill=cOnwind, draw=none, only no, forget plot] table[x=Year, y={Onshore wind}] {\capsAll};
\addplot[bar shift=4pt, fill=cOffwind, draw=none, only no, forget plot] table[x=Year, y={Offshore wind}] {\capsAll};
\addplot[bar shift=4pt, fill=cSolarFixed,  draw=none, only no, forget plot] table[x=Year, y={Solar fixed tilt}] {\capsAll};
\addplot[bar shift=4pt, fill=cSolarHSAT, draw=none, only no, forget plot] table[x=Year, y={Solar HSAT}] {\capsAll};
\addplot[bar shift=4pt, fill=cSolarRooftop, draw=none, only no, forget plot] table[x=Year, y={Solar rooftop}] {\capsAll};

% =================================================================
% Sweden
% =================================================================
\nextgroupplot[
  xlabel={Sweden},
  axis y line=left, y axis line style={opacity=0}, ytick style={draw=none},
]
\addplot[bar shift=-4pt, fill=cNuclear, draw=none, forget plot, postaction={pattern=north east lines, pattern color=black!20}] table[x=Year, y=Nuclear] {\capSEngdp};
\addplot[bar shift=-4pt, fill=cHydro,  draw=none, forget plot, postaction={pattern=north east lines, pattern color=black!20}] table[x=Year, y=Hydro] {\capSEngdp};
\addplot[bar shift=-4pt, fill=cOnwind, draw=none, forget plot, postaction={pattern=north east lines, pattern color=black!20}] table[x=Year, y=WindOnshore] {\capSEngdp};
\addplot[bar shift=-4pt, fill=cOffwind, draw=none, forget plot, postaction={pattern=north east lines, pattern color=black!20}] table[x=Year, y=WindOffshore] {\capSEngdp};
\addplot[bar shift=-4pt, fill=cSolar,  draw=none, forget plot, postaction={pattern=north east lines, pattern color=black!20}] table[x=Year, y=Solar] {\capSEngdp};
\addplot[bar shift=-4pt, fill=cThermal, draw=none, forget plot, postaction={pattern=north east lines, pattern color=black!20}] table[x=Year, y=Thermal] {\capSEngdp};
\resetstackedplots
\addplot[bar shift=4pt, fill=cNuclear, draw=none, only se, forget plot] table[x=Year, y={Nuclear}] {\capsAll};
\addplot[bar shift=4pt, fill=cHydro, draw=none, only se, forget plot] table[x=Year, y={Hydroelectricity}] {\capsAll};
\addplot[bar shift=4pt, fill=cOnwind, draw=none, only se, forget plot] table[x=Year, y={Onshore wind}] {\capsAll};
\addplot[bar shift=4pt, fill=cOffwind, draw=none, only se, forget plot] table[x=Year, y={Offshore wind}] {\capsAll};
\addplot[bar shift=4pt, fill=cSolarFixed,  draw=none, only se, forget plot] table[x=Year, y={Solar fixed tilt}] {\capsAll};
\addplot[bar shift=4pt, fill=cSolarHSAT, draw=none, only se, forget plot] table[x=Year, y={Solar HSAT}] {\capsAll};
\addplot[bar shift=4pt, fill=cSolarRooftop, draw=none, only se, forget plot] table[x=Year, y={Solar rooftop}] {\capsAll};

\end{groupplot}

% --- shared legend below the groupplot ---
\node[anchor=north, yshift=-0.2cm]
  at (current bounding box.south) {\pgfplotslegendfromname{combinedcaplegend}};

\end{tikzpicture}
  \vspace*{-0.7cm}
\caption{Installed generation capacity per Nordic country.
The NGDP projections, displayed with hatched bars, are shown from 2025, 2030, 2040, and 2050, as the left stacked bars.
The PyPSA-Eur simulation results, in 5-year increments, as the right stacked bars.}\label{fig:1}
  \vspace*{-0.1cm}
\end{figure*}

To this effect, this work explores a least-cost pathway for the Nordic power system in 2050, under a net-zero constraint and explicit network representation.
We take the basis from the Nordic Development Plan 2025 (NGDP)~\cite{NGDP2025}, which provides an informed set of predictions, drawn up by the Nordics transmission system operators (TSOs), of the expected evolution of the Nordic system's needs and targets, informed by their respective distribution operators, national authorities and governmental bodies, and the resulting implications for grid development.
Under a strict evaluation of the targets of the NGDP, we compare the described development of solar and wind capacity expansion, as well as the implementation of hydrogen electrolysis.
We employ PyPSA-Eur to quantify the transition pathways from 2025 to 2050, in 5-year steps, and explore the resulting difference to the NGDP targets~\cite{Hoersch2018,Brown2018,Saleem2024,Malum2025}.
In this regard, we explore the implications of the NGDP targets for the Nordic system's balance, and the role of different technologies in meeting the demand growth and net-zero target, as well as how a least-cost pathway differs from the targets and expected development of the NGDP.

\section{Methods}

\subsection{Model framework and optimisation problem}
We model the Nordic electricity system with the open-source capacity-expansion and dispatch framework PyPSA-Eur (post-version v2025.07.0, January 2026)~\cite{Brown2018,Hoersch2018,Neumann2023}.
The model co-optimises investment in generation, storage, and transmission with system operation over a full weather year, subject to nodal power balance, network constraints, and technology-specific operating limits.
The optimisation target is to minimise total annual system cost under a strict \ce{CO2} budget, comprising annualised investment costs and operating costs.

% To reduce computational complexity while retaining the principal transmission topology, the high-voltage network is clustered into 100 representative buses. PyPSA-Eur aggregates substations and lines using k-means clustering, and assigns renewable potentials to clustered regions via Voronoi cells. The transmission grid is represented as an equivalent \qty{380}{kV} network in which all aggregated lines are assumed to operate at the same voltage level.

The model uses a standard PyPSA-Eur data pipeline, combining grid infrastructure, power plant inventories, meteorological reanalysis, and land-use constraints (a more detailed setup can be found in the original Refs.~\cite{Brown2018,Hoersch2018,Neumann2023})
The transmission network topology is constructed from ENTSO-E map data using OpenStreetMaps for geospatial referencing of the transmission elements.
Conventional power plants and existing generation capacities are assembled and harmonised with existing capacity statistics.
Renewable energy availability is derived from ERA5 reanalysis data, converting meteorological fields, notably wind speeds and solar irradiance, into spatially resolved capacity-factor time series for wind and solar technologies at each bus region.
From this, land eligibility for onshore renewables is masked based on CORINE land cover and Natura 2000 protected areas to assert spatial constraints on deployment\footnote{Norway is not part of the Natura 2000 network, and thus the corresponding exclusions cannot be applied within the same dataset framework. This has only a minute impact in the simulations, given the weak expansion of onshore wind power in Norway.}.

PyPSA and PyPSA-Eur solve a large mixed-integer linear programming problem, considering the investment and operational decisions of generation, storage, and transmission assets, subject to a set of physical constraints, obeying Kirchhoff's laws, and a sequential reduction of \ce{CO2} emissions by year, culminating in a net-zero \ce{CO2} emissions in 2050.
Particularly important for this model are the assumptions on the capacity density of solar and wind power generation technologies.
Herein, we consider a maximum capacity density of \qty{6.2}{MW/km^2} for onshore wind, \qty{5.5}{MW/km^2} for offshore wind, and \qty{43}{MW/km^2} for solar PV technologies, both utility PV fixed tilt and rooftop, and \qty{43}{MW/km^2} for solar PV with horizontal single-axis tracking (HSAT)~\cite{Borrmann2018,Covey2026}.

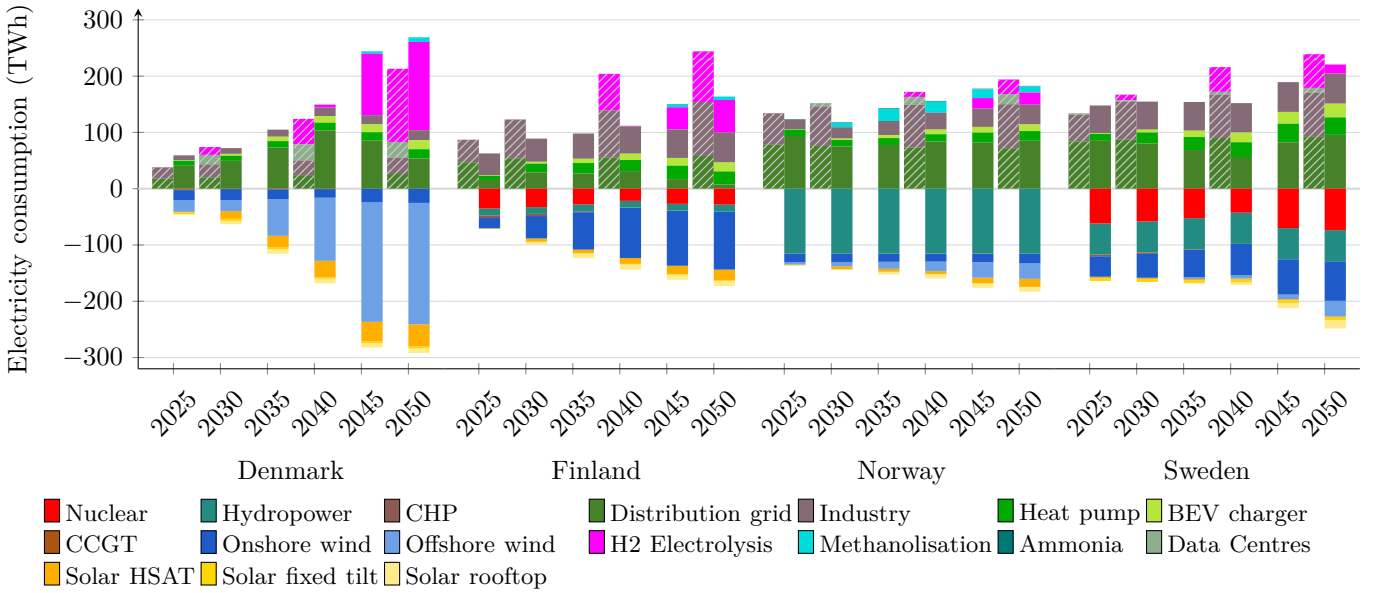
\begin{figure*}[ht]
\centering
\begin{tikzpicture}

\pgfplotsset{
  mergedbalance/.style={
    ybar stacked,
    stack negative=separate,
    bar width=8pt,
    symbolic x coords={2025,2030,2035,2040,2045,2050},
    xtick=data,
    x tick label style={rotate=45},
    ymin=-320, ymax=320,
    ytick={-300,-200,-100,0,100,200,300},
    ymajorgrids,
    major grid style={black!15},
    axis x line=bottom,
    axis y line=left,
    enlarge x limits=0.15,
    extra y ticks={0},
    extra y tick labels={},
    extra y tick style={
 grid=major,
 grid style={black, thick},
 tick style={draw=none},
    },
  }
}

\begin{groupplot}[
  group style={
    group size=4 by 1,
    horizontal sep=0cm,
    ylabels at=edge left,
    yticklabels at=edge left,
  },
  width=0.31\linewidth,
  height=0.35\linewidth,
  mergedbalance,
  x axis line style={-},
]

% ==========================================================
% Denmark
% ==========================================================
\nextgroupplot[
  xlabel={Denmark},
  ylabel={Electricity consumption (\unit{TWh})},
  inner frame sep=0pt
]
% --- PyPSA bars (shifted right, solid) ---
\addplot[bar shift=4pt, fill=cNuclear, draw=none, only dk] table[x=Year, y expr=-\thisrowno{2}]  {\balancesAll};
\addplot[bar shift=4pt, fill=cHydro, draw=none, only dk] table[x=Year, y expr=-\thisrowno{3}]  {\balancesAll};
\addplot[bar shift=4pt, fill=cCHP, draw=none, only dk] table[x=Year, y expr=-\thisrowno{4}]  {\balancesAll};
\addplot[bar shift=4pt, fill=cCCGT,  draw=none, only dk] table[x=Year, y expr=-\thisrowno{5}]  {\balancesAll};
\addplot[bar shift=4pt, fill=cOnwind, draw=none, only dk] table[x=Year, y expr=-\thisrowno{6}]  {\balancesAll};
\addplot[bar shift=4pt, fill=cOffwind, draw=none, only dk] table[x=Year, y expr=-\thisrowno{7}]  {\balancesAll};
\addplot[bar shift=4pt, fill=cSolarHSAT,  draw=none, only dk] table[x=Year, y expr=-\thisrowno{8}]  {\balancesAll};
\addplot[bar shift=4pt, fill=cSolarFixed,  draw=none, only dk] table[x=Year, y expr=-\thisrowno{9}]  {\balancesAll};
\addplot[bar shift=4pt, fill=cSolarRooftop, draw=none, only dk] table[x=Year, y expr=-\thisrowno{10}]  {\balancesAll};
\addplot[bar shift=4pt, fill=cDist, draw=none, only dk] table[x=Year, y expr=-\thisrowno{11}] {\balancesAll};
\addplot[bar shift=4pt, fill=cHeatPump, draw=none, only dk] table[x=Year, y expr=-\thisrowno{13}] {\balancesAll};
\addplot[bar shift=4pt, fill=cBEV, draw=none, only dk] table[x=Year, y expr=-\thisrowno{14}] {\balancesAll};
\addplot[bar shift=4pt, fill=cIndustry, draw=none, only dk] table[x=Year, y expr=-\thisrowno{12}] {\balancesAll};
\addplot[bar shift=4pt, fill=cHydrogen, draw=none, only dk] table[x=Year, y expr=-\thisrowno{15}] {\balancesAll};
\addplot[bar shift=4pt, fill=cMeth, draw=none, only dk] table[x=Year, y expr=-\thisrowno{16}] {\balancesAll};
\addplot[bar shift=4pt, fill=cAmmo, draw=none, only dk] table[x=Year, y expr=-\thisrowno{17}] {\balancesAll};
%\addplot[bar shift=4pt, fill=cExport,  draw=none, only dk] table[x=Year, y expr=-\thisrowno{18}] {\balancesAll};
\resetstackedplots
% --- NGDP bars (shifted left, hatched) ---
\addplot[bar shift=-4pt, fill=cDist, draw=none, postaction={pattern=north east lines, pattern color=black!20}] table[x=Year, y=Households] {\consDKngdp};
\addplot[bar shift=-4pt, fill=cIndustry,  draw=none, postaction={pattern=north east lines, pattern color=black!20}] table[x=Year, y=Industry]   {\consDKngdp};
%\addplot[bar shift=-4pt, fill=cTransport, draw=none, postaction={pattern=north east lines, pattern color=black!20}] table[x=Year, y=Transport]  {\consDKngdp};
\addplot[bar shift=-4pt, fill=cDataCentre, draw=none, postaction={pattern=north east lines, pattern color=black!20}] table[x=Year, y=DataCentre] {\consDKngdp};
\addplot[bar shift=-4pt, fill=cHydrogen,  draw=none, postaction={pattern=north east lines, pattern color=black!20}] table[x=Year, y=Hydrogen]   {\consDKngdp};

% ==========================================================
% Finland
% ==========================================================
\nextgroupplot[
  xlabel={Finland},
  axis y line=left, y axis line style={opacity=0}, ytick style={draw=none},
]
% PyPSA bars (shifted right, solid)
\addplot[bar shift=4pt, fill=cNuclear, draw=none, only fi, forget plot] table[x=Year, y expr=-\thisrowno{2}]  {\balancesAll};
\addplot[bar shift=4pt, fill=cHydro, draw=none, only fi, forget plot] table[x=Year, y expr=-\thisrowno{3}]  {\balancesAll};
\addplot[bar shift=4pt, fill=cCHP, draw=none, only fi, forget plot] table[x=Year, y expr=-\thisrowno{4}]  {\balancesAll};
\addplot[bar shift=4pt, fill=cCCGT,  draw=none, only fi, forget plot] table[x=Year, y expr=-\thisrowno{5}]  {\balancesAll};
\addplot[bar shift=4pt, fill=cOnwind, draw=none, only fi, forget plot] table[x=Year, y expr=-\thisrowno{6}]  {\balancesAll};
\addplot[bar shift=4pt, fill=cOffwind, draw=none, only fi, forget plot] table[x=Year, y expr=-\thisrowno{7}]  {\balancesAll};
\addplot[bar shift=4pt, fill=cSolarHSAT,  draw=none, only fi, forget plot] table[x=Year, y expr=-\thisrowno{8}]  {\balancesAll};
\addplot[bar shift=4pt, fill=cSolarFixed,  draw=none, only fi, forget plot] table[x=Year, y expr=-\thisrowno{9}]  {\balancesAll};
\addplot[bar shift=4pt, fill=cSolarRooftop, draw=none, only fi, forget plot] table[x=Year, y expr=-\thisrowno{10}]  {\balancesAll};
\addplot[bar shift=4pt, fill=cDist, draw=none, only fi, forget plot] table[x=Year, y expr=-\thisrowno{11}] {\balancesAll};
\addplot[bar shift=4pt, fill=cHeatPump, draw=none, only fi, forget plot] table[x=Year, y expr=-\thisrowno{13}] {\balancesAll};
\addplot[bar shift=4pt, fill=cBEV, draw=none, only fi, forget plot] table[x=Year, y expr=-\thisrowno{14}] {\balancesAll};
\addplot[bar shift=4pt, fill=cIndustry, draw=none, only fi, forget plot] table[x=Year, y expr=-\thisrowno{12}] {\balancesAll};
\addplot[bar shift=4pt, fill=cHydrogen, draw=none, only fi, forget plot] table[x=Year, y expr=-\thisrowno{15}] {\balancesAll};
\addplot[bar shift=4pt, fill=cMeth, draw=none, only fi, forget plot] table[x=Year, y expr=-\thisrowno{16}] {\balancesAll};
\addplot[bar shift=4pt, fill=cAmmo, draw=none, only fi, forget plot] table[x=Year, y expr=-\thisrowno{17}] {\balancesAll};
%\addplot[bar shift=4pt, fill=cExport,  draw=none, only fi, forget plot] table[x=Year, y expr=-\thisrowno{18}] {\balancesAll};
\resetstackedplots
% NGDP bars (shifted left, hatched)
\addplot[bar shift=-4pt, fill=cDist, draw=none, forget plot, postaction={pattern=north east lines, pattern color=black!20}] table[x=Year, y=Households] {\consFIngdp};
\addplot[bar shift=-4pt, fill=cIndustry,  draw=none, forget plot, postaction={pattern=north east lines, pattern color=black!20}] table[x=Year, y=Industry]   {\consFIngdp};
%\addplot[bar shift=-4pt, fill=cTransport, draw=none, forget plot, postaction={pattern=north east lines, pattern color=black!20}] table[x=Year, y=Transport]  {\consFIngdp};
\addplot[bar shift=-4pt, fill=cDataCentre, draw=none, forget plot, postaction={pattern=north east lines, pattern color=black!20}] table[x=Year, y=DataCentre] {\consFIngdp};
\addplot[bar shift=-4pt, fill=cHydrogen,  draw=none, forget plot, postaction={pattern=north east lines, pattern color=black!20}] table[x=Year, y=Hydrogen]   {\consFIngdp};

% ==========================================================
% Norway
% ==========================================================
\nextgroupplot[
  xlabel={Norway},
  axis y line=left, y axis line style={opacity=0}, ytick style={draw=none},
]
% PyPSA bars (shifted right, solid)
\addplot[bar shift=4pt, fill=cNuclear, draw=none, only no, forget plot] table[x=Year, y expr=-\thisrowno{2}]  {\balancesAll};
\addplot[bar shift=4pt, fill=cHydro, draw=none, only no, forget plot] table[x=Year, y expr=-\thisrowno{3}]  {\balancesAll};
\addplot[bar shift=4pt, fill=cCHP, draw=none, only no, forget plot] table[x=Year, y expr=-\thisrowno{4}]  {\balancesAll};
\addplot[bar shift=4pt, fill=cCCGT,  draw=none, only no, forget plot] table[x=Year, y expr=-\thisrowno{5}]  {\balancesAll};
\addplot[bar shift=4pt, fill=cOnwind, draw=none, only no, forget plot] table[x=Year, y expr=-\thisrowno{6}]  {\balancesAll};
\addplot[bar shift=4pt, fill=cOffwind, draw=none, only no, forget plot] table[x=Year, y expr=-\thisrowno{7}]  {\balancesAll};
\addplot[bar shift=4pt, fill=cSolarHSAT,  draw=none, only no, forget plot] table[x=Year, y expr=-\thisrowno{8}]  {\balancesAll};
\addplot[bar shift=4pt, fill=cSolarFixed,  draw=none, only no, forget plot] table[x=Year, y expr=-\thisrowno{9}]  {\balancesAll};
\addplot[bar shift=4pt, fill=cSolarRooftop, draw=none, only no, forget plot] table[x=Year, y expr=-\thisrowno{10}]  {\balancesAll};
\addplot[bar shift=4pt, fill=cDist, draw=none, only no, forget plot] table[x=Year, y expr=-\thisrowno{11}] {\balancesAll};
\addplot[bar shift=4pt, fill=cHeatPump, draw=none, only no, forget plot] table[x=Year, y expr=-\thisrowno{13}] {\balancesAll};
\addplot[bar shift=4pt, fill=cBEV, draw=none, only no, forget plot] table[x=Year, y expr=-\thisrowno{14}] {\balancesAll};
\addplot[bar shift=4pt, fill=cIndustry, draw=none, only no, forget plot] table[x=Year, y expr=-\thisrowno{12}] {\balancesAll};
\addplot[bar shift=4pt, fill=cHydrogen, draw=none, only no, forget plot] table[x=Year, y expr=-\thisrowno{15}] {\balancesAll};
\addplot[bar shift=4pt, fill=cMeth, draw=none, only no, forget plot] table[x=Year, y expr=-\thisrowno{16}] {\balancesAll};
\addplot[bar shift=4pt, fill=cAmmo, draw=none, only no, forget plot] table[x=Year, y expr=-\thisrowno{17}] {\balancesAll};
%\addplot[bar shift=4pt, fill=cExport,  draw=none, only no, forget plot] table[x=Year, y expr=-\thisrowno{18}] {\balancesAll};
\resetstackedplots
% NGDP bars (shifted left, hatched)
\addplot[bar shift=-4pt, fill=cDist, draw=none, forget plot, postaction={pattern=north east lines, pattern color=black!20}] table[x=Year, y=Households] {\consNOngdp};
\addplot[bar shift=-4pt, fill=cIndustry,  draw=none, forget plot, postaction={pattern=north east lines, pattern color=black!20}] table[x=Year, y=Industry]   {\consNOngdp};
%\addplot[bar shift=-4pt, fill=cTransport, draw=none, forget plot, postaction={pattern=north east lines, pattern color=black!20}] table[x=Year, y=Transport]  {\consNOngdp};
\addplot[bar shift=-4pt, fill=cDataCentre, draw=none, forget plot, postaction={pattern=north east lines, pattern color=black!20}] table[x=Year, y=DataCentre] {\consNOngdp};
\addplot[bar shift=-4pt, fill=cHydrogen,  draw=none, forget plot, postaction={pattern=north east lines, pattern color=black!20}] table[x=Year, y=Hydrogen]   {\consNOngdp};

% ==========================================================
% Sweden
% ==========================================================
\nextgroupplot[
  xlabel={Sweden},
  axis y line=left, y axis line style={opacity=0}, ytick style={draw=none},
]
% PyPSA bars (shifted right, solid)
\addplot[bar shift=4pt, fill=cNuclear, draw=none, only se, forget plot] table[x=Year, y expr=-\thisrowno{2}]  {\balancesAll};
\addplot[bar shift=4pt, fill=cHydro, draw=none, only se, forget plot] table[x=Year, y expr=-\thisrowno{3}]  {\balancesAll};
\addplot[bar shift=4pt, fill=cCHP, draw=none, only se, forget plot] table[x=Year, y expr=-\thisrowno{4}]  {\balancesAll};
\addplot[bar shift=4pt, fill=cCCGT, draw=none, only se, forget plot] table[x=Year, y expr=-\thisrowno{5}]  {\balancesAll};
\addplot[bar shift=4pt, fill=cOnwind, draw=none, only se, forget plot] table[x=Year, y expr=-\thisrowno{6}]  {\balancesAll};
\addplot[bar shift=4pt, fill=cOffwind, draw=none, only se, forget plot] table[x=Year, y expr=-\thisrowno{7}]  {\balancesAll};
\addplot[bar shift=4pt, fill=cSolarHSAT, draw=none, only se, forget plot] table[x=Year, y expr=-\thisrowno{8}]  {\balancesAll};
\addplot[bar shift=4pt, fill=cSolarFixed, draw=none, only se, forget plot] table[x=Year, y expr=-\thisrowno{9}]  {\balancesAll};
\addplot[bar shift=4pt, fill=cSolarRooftop, draw=none, only se, forget plot] table[x=Year, y expr=-\thisrowno{10}]  {\balancesAll};
\addplot[bar shift=4pt, fill=cDist, draw=none, only se, forget plot] table[x=Year, y expr=-\thisrowno{11}] {\balancesAll};
\addplot[bar shift=4pt, fill=cHeatPump, draw=none, only se, forget plot] table[x=Year, y expr=-\thisrowno{13}] {\balancesAll};
\addplot[bar shift=4pt, fill=cBEV, draw=none, only se, forget plot] table[x=Year, y expr=-\thisrowno{14}] {\balancesAll};
\addplot[bar shift=4pt, fill=cIndustry, draw=none, only se, forget plot] table[x=Year, y expr=-\thisrowno{12}] {\balancesAll};
\addplot[bar shift=4pt, fill=cHydrogen, draw=none, only se, forget plot] table[x=Year, y expr=-\thisrowno{15}] {\balancesAll};
\addplot[bar shift=4pt, fill=cMeth, draw=none, only se, forget plot] table[x=Year, y expr=-\thisrowno{16}] {\balancesAll};
\addplot[bar shift=4pt, fill=cAmmo, draw=none, only se, forget plot] table[x=Year, y expr=-\thisrowno{17}] {\balancesAll};
%\addplot[bar shift=4pt, fill=cExport, draw=none, only se, forget plot] table[x=Year, y expr=-\thisrowno{18}] {\balancesAll};
\resetstackedplots
% NGDP bars (shifted left, hatched)
\addplot[bar shift=-4pt, fill=cDist, draw=none, forget plot, postaction={pattern=north east lines, pattern color=black!20}] table[x=Year, y=Households] {\consSEngdp};
\addplot[bar shift=-4pt, fill=cIndustry,  draw=none, forget plot, postaction={pattern=north east lines, pattern color=black!20}] table[x=Year, y=Industry]   {\consSEngdp};
%\addplot[bar shift=-4pt, fill=cTransport, draw=none, forget plot, postaction={pattern=north east lines, pattern color=black!20}] table[x=Year, y=Transport]  {\consSEngdp};
\addplot[bar shift=-4pt, fill=cDataCentre, draw=none, forget plot, postaction={pattern=north east lines, pattern color=black!20}] table[x=Year, y=DataCentre] {\consSEngdp};
\addplot[bar shift=-4pt, fill=cHydrogen,  draw=none, forget plot, postaction={pattern=north east lines, pattern color=black!20}] table[x=Year, y=Hydrogen]   {\consSEngdp};

\end{groupplot}

% --- Generation legend (left, below) ---
\begin{axis}[
  hide axis,
  width=0pt, height=0pt,
  scale only axis,
  xmin=0, xmax=1, ymin=0, ymax=1,
  at={(current bounding box.south west)},
  anchor=north west,
  yshift=0cm,
  xshift=0.5cm,
  legend image code/.code={%
    \draw[fill=#1] (0cm,-0.1cm) rectangle (0.2cm,0.2cm);},
  legend columns=3,
  legend cell align=left,
  legend style={
 draw=none, fill=none, font=\small,
    anchor=north west,
  },
]
\addlegendimage{ fill=cNuclear, draw=none}     \addlegendentry{Nuclear}
\addlegendimage{fill=cHydro, draw=none}  \addlegendentry{Hydropower}
\addlegendimage{fill=cCHP, draw=none}    \addlegendentry{CHP}
\addlegendimage{fill=cCCGT, draw=none}   \addlegendentry{CCGT}
\addlegendimage{fill=cOnwind, draw=none} \addlegendentry{Onshore wind}
\addlegendimage{fill=cOffwind, draw=none} \addlegendentry{Offshore wind}
\addlegendimage{fill=cSolarHSAT, draw=none}    \addlegendentry{Solar HSAT}
\addlegendimage{fill=cSolarFixed, draw=none}   \addlegendentry{Solar fixed tilt}
\addlegendimage{fill=cSolarRooftop, draw=none} \addlegendentry{Solar rooftop}
\end{axis}

% --- Demand legend (right, below) ---
\begin{axis}[
  hide axis,
  width=0pt, height=0pt,
  scale only axis,
  xmin=0, xmax=1, ymin=0, ymax=1,
  at={(current bounding box.south east)},
  anchor=north east,
  yshift=1.4cm,
  xshift=-0.5cm,
  legend image code/.code={%
    \draw[fill=#1] (0cm,-0.1cm) rectangle (0.2cm,0.2cm);},
  legend columns=4,
  legend cell align=left,
  inner sep=2cm,
  legend style={
 draw=none, fill=none, font=\small,
    anchor=north east,
  },
]
\addlegendimage{ fill=cDist, draw=none}     \addlegendentry{Distribution grid}
\addlegendimage{ fill=cIndustry, draw=none} \addlegendentry{Industry}
\addlegendimage{ fill=cHeatPump, draw=none} \addlegendentry{Heat pump}
\addlegendimage{ fill=cBEV, draw=none} \addlegendentry{BEV charger}
\addlegendimage{ fill=cHydrogen, draw=none} \addlegendentry{H2 Electrolysis}
\addlegendimage{ fill=cMeth, draw=none}     \addlegendentry{Methanolisation}
\addlegendimage{ fill=cAmmo, draw=none}     \addlegendentry{Ammonia}
\addlegendimage{ fill=cDataCentre, draw=none}     \addlegendentry{Data Centres}
%\addlegendimage{area legend, fill=cExport, draw=none}   \addlegendentry{Power export}
\end{axis}

\end{tikzpicture}
  \vspace*{-0.1cm}
\caption{Per-country energy balance from 2025 to 2050.
The NGDP projections, displayed with hatched bars, are shown from 2025, 2030, 2040, and 2050, as the left stacked bars.
The PyPSA-Eur simulation results, in 5-year increments, as the right stacked bars.
The consumption targets follow PyPSA-Eur's targets for population density increase and GDP growth, and disagree strongly with the Finnish predictions, particularly regarding the development of industry in the country, and the Danish distribution-level domain, speaking to a mismatch in the manner in which electricity and heating are accounted for in Denmark.}\label{fig:2}
\vspace*{-0.1cm}
\end{figure*}

\subsection{Technology representation and constraints}
The system includes conventional generators (e.g. gas and nuclear), renewable generators (onshore, offshore, and floating offshore wind, run-of-river, utility solar PV, solar rooftop photovoltaics (PV), and horizontal single-axis tracking PV), and energy storage technologies (hydropower reservoirs, pumped hydro storage, battery, hydrogen, and methanolisation).
The optimisation allows all technologies to be built up.
Nuclear, solar, wind, and hydropower generators are subject to a lower-bound and upper-bound capacity constraint dictated by the NGDP, meaning they are bound to the lowest capacity of the previous decade and the highest of the following decade, respectively.
Non-renewable technologies are free to be built up as the NGDP places no constraints on their capacity\footnote{Non-renewable technologies are phased out quickly due to high costs and \ce{CO2} budget constraints.}.
Noteworthy is that the simulation does not assume an expansion of demand following the NGDP targets, but rather follows a demand rise following estimates for population and GDP growth, affecting both demand for agriculture, industry, and distribution-level consumption such as households.
Transport (except for electric vehicles) and data centre demand are not modelled.

\subsection{NGDP scenario design and modelling of the North Sea-centric European electricity system}
To ensure the model appropriately reflects the targets of a renewable Nordic power system pathway, all European countries that have or will have a substantial stake in the North Sea offshore wind development are included in the model domain.
Moreover, due to the importance of the Baltics and Poland for the Nordic system's interconnections, they are also included in the model domain, thus, Great Britain, Northern Ireland, the Netherlands, Belgium, France, Germany, Estonia, Latvia, Lithuania, and Poland are included in the model domain, but each is spatially aggregated to a single bus (not shown in the figures).

The NGDP provides predictions for 2025, 2030, 2040, and 2050, but not for the intermediate years.
We evaluate a pathway scenario based on the NGDP for 5-year intervals from 2025 to 2050.
This includes Finland, Sweden, Norway, and Denmark, which we denote as a whole as the Nordics.
The NGDP provides a comprehensive set of targets for the Nordic system, including the expected installed capacities by technology, the required transmission reinforcements, and the anticipated demand growth.

\section{Results and discussion}

In order to mimic the explicit targets of the predicted installed capacity, we track the generation from the NGDP in the model.
In Fig.~\ref{fig:1}, we show the targeted capacities extracted from the NGDP alongside the PyPSA-Eur model.
We note, firstly, that we do not account for most thermal generation in the Nordics, but generally, solar generation (comprising fixed-tilt PV system, HSAT, and rooftop technologies), offshore wind (comprising AC, DC, and floating technologies), onshore wind, hydropower (comprising hydroelectricity with catchment and run-of river), generally match the targets of the NGDP.
The primary discrepancies arise from the fact that, in our modelling, the targets for each 5-year prediction accept a range of installed capacities within the bounds of the previous and following period (e.g., generation capacity for 2035 is bound between the NGDP targets from 2030 and 2040).
Smaller adjustments are applied for the nuclear generation in Sweden to match the countries targets.
The largest discrepancies in the model amount to inconsistencies between hydropower generation in Norway, which is difficult to accurately account for due to the very distributed nature, where over \num{2000} known generating units are known.

Overall, we note that, under similar consumption assumptions for the Nordics (see Fig.~\ref{fig:2}), the generation capacities match the NGDP.
One key point to note is the differentiation of the solar generation.
Following our model, Dermark invests heavily in large HSAT solar systems from the onset in 2030, whereas solar rooftop technologies play a larger role in Finland, Norway, and Sweden, accounting for \qty{30}{\percent} or more of the overall installed capacities.

In Fig.~\ref{fig:2}, the electricity consumption and generation from the NGDP and the simulation are shown.
The positive-valued elements represent the consumption by technologies or asset type.
The negative-valued elements represent the energy generation.
The distribution grid accounts for all consumption from household energy in the form of electricity and heating.
Heat pumps and electric vehicles (BEV) are simulated separately.
NGDP includes data centres as predicted future demand, which are not accounted for in PyPSA-Eur.

\begin{figure}[t]
  \includegraphics[width=\linewidth]{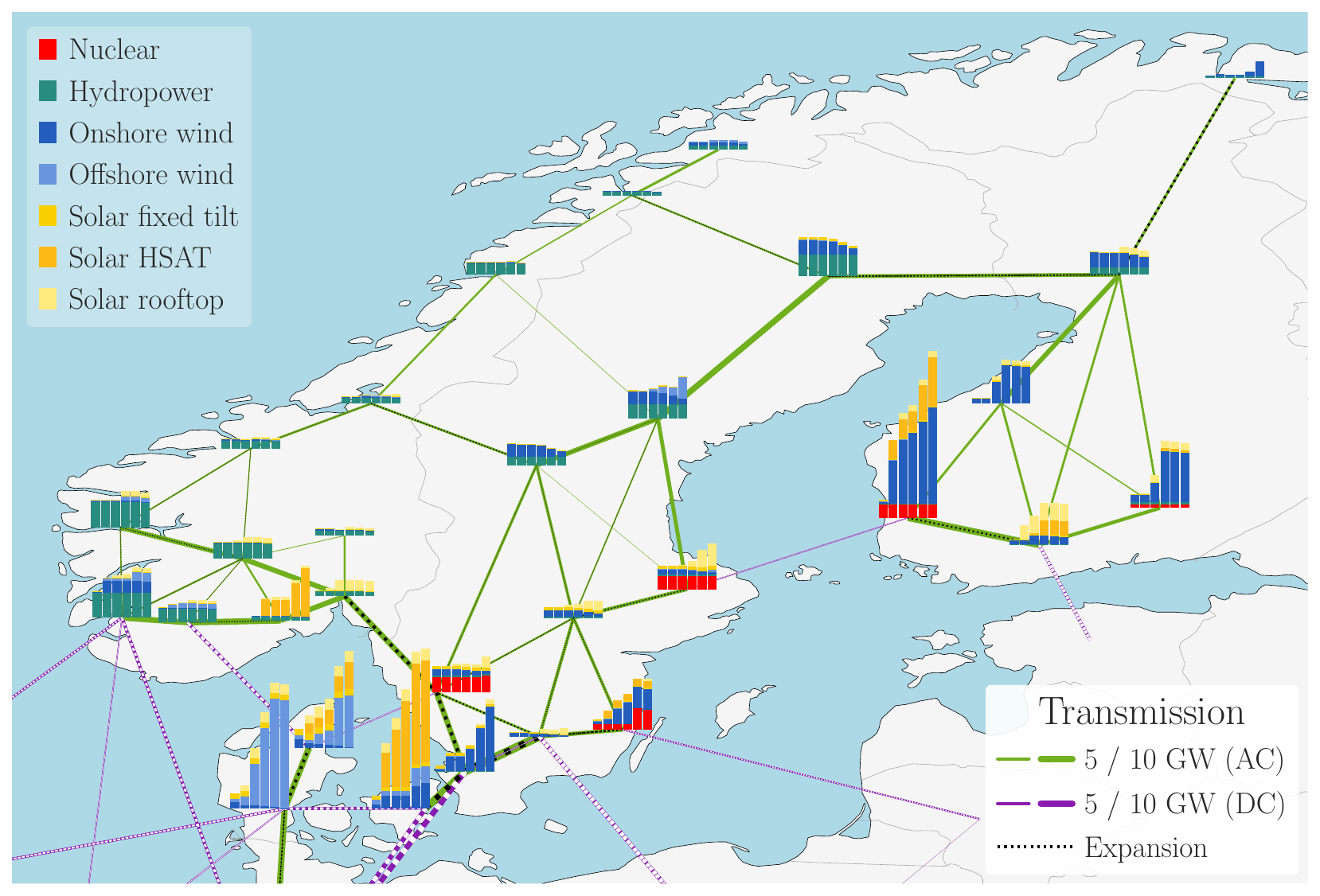}
  \vspace*{-0.5cm}
  \caption{Geographical distribution of generation capacity expansion in the PyPSA-Eur optimised scenario.
  The stacked bars indicate the installed capacities per 5-year scenario from 2025 to 2050.
  A large solar and offshore wind expansion in Denmark, along with a large onshore wind expansion in Finland, constitute the largest changes in the generation capacities in the Nordic system.
  All HVDC connections with Continental Europe are reinforced. 
  }\label{fig:3}
  \vspace*{-0.3cm}
\end{figure}

Denmark's distribution grid, i.e., household energy demand, is expected to grow little following the NGDP, which contrasts strongly with the expected grow simulated by PyPSA-Eur.
For Finland, Norway, and Sweden, the expectations of the demand at the distribution level match adequately.
The fairly contrasting aspect within all simulations is the disagreement between the expected growth of industry simulated by PyPSA-Eur.
A clear case is the estimation that Finnish industry's electrical needs will more than double between 2025 and 2050, according to the NGDP, moving from \qty{40}{TWh} in 2025 to \qty{96}{TWh} in 2050.
The expected growth simulated by PyPSA-Eur increases from \qty{38.91}{TWh} in 2025 to \qty{53.08}{TWh} in 2050.
We find similar indications for Norway, wherein the NGDP points to an increase of from \qty{56}{TWh} in 2025 to \qty{78}{TWh} in 2050, whereas PyPSA-Eur estimates \qty{17.25}{TWh} in 2025 to \qty{35.19}{TWh} in 2050, which are very shy numbers in contrast.
Here we should note that PyPSA-Eur consider the gas and oil industry a separate element to general industry, and thus the gap in demand for industry in Norway is likely due to this separation.
With this in mind, our simulations see a doubling of energy consumption for non-gas-and-oil in Norway, which aligns with the countries targets.
Whether a doubling of the industry demand can be expected in Finland, this is up to debate.

Hydrogen technologies play a markedly different role in the two scenarios: while the NGDP projects early and steadily increasing hydrogen demand, starting already in 2030 in Denmark and Sweden, PyPSA-Eur estimates a much later hydrogen deployment ramping up only in 2045 and 2050, primarily as a flexibility and seasonal storage solution in a highly VRE-dominated system.
This shifts the need and use of hydrogen technologies by a decade due primarily to a need for deployment of a large-scale VRE to ensure excess energy can be stored, while simultaneously ensuring hydrogen technology costs have sufficiently reduced to make them economically attractive.
In agreement with the NGDP, PyPSA-Eur projects Denmark as the prime candidate for a strong expansion of offshore wind and solar power coupled with a hydrogen economy.
Finland's bet on a similar hydrogen based economy is also present.
In contrast, investments in hydrogen technology in Sweden come late and are scarce.
Norway's hydrogen targets generally align with the NGDP, yet methanolisation is preferred over hydrogen and is explored from 2030 onward.

In Fig.~\ref{fig:3}, the geographical distribution of the generation capacity expansion in the PyPSA-Eur optimised scenario is shown, where the stacked bars indicate the installed capacities per 5-year scenario from 2025 to 2050.
A large solar and offshore wind expansion in Denmark, along with a large onshore wind expansion in Finland, constitute the largest changes in the generation capacities in the Nordic system.
We note that in Norway, the expansion of the technologies is very centralised, seeing all substantial expansion in solar HSAT technologies south west of Oslo, which is the area with the highest solar potential in the country.
Sweden's expansion distributed practically in the southern half of the country, and Finland's investment in onshore wind is fairly distributed, but mainly concentrated in the south west of the country.
Denmark, the country with the largest expansion of offshore wind and solar power, is simulated with a low bus density (only three buses), which does not allow for a more detailed geographical distribution of the generation expansion.
High-voltage (HV) AC connections are mainly upgraded in the south of Sweden and with the general aim to transporting power to and from Continental Europe via Denmark.
This contrasts strongly with the TSO targets that report a very large investment of the reinforcement of the internal transmission grids.
This, we note, is a limitation of our model, which does not allow for a detailed representation of the internal transmission grid, casting all internal transmission level voltage to a singular high-voltage, thus missing the potential for the targetted reinforcements in Norway in with the increase of the voltage levels from \qty{300}{kV} to \qty{420}{kV}.
The NGDP details various targets for extensive grid upgrades in the next 5 to 10 years, which are not captured in our model.
On the other hand, all HVDC connections with Continental Europe and between the Nordic countries are strongly reinforced, which is in line with the NGDP's targets for the interconnections of the Nordic system.
Here, we note that capital expenditure values separating these are not available in the NGDP and thus hard to contrast with the model's results, but the narrative of the models aligns.

\begin{figure}[t]
\hspace*{-0.1cm}\begin{tikzpicture}
\pgfplotsset{
  sidebysidecap/.style={
    ybar stacked,
    bar width=7pt,
    symbolic x coords={2025,2030,2035,2040,2045,2050},
    xtick=data,
    xmin={2030},
    x tick label style={rotate=90},
    x axis line style={-},
    ymin=0, ymax=16,
    ytick={0,3,6,9,12,15},
    ymajorgrids,
    major grid style={black!15},
    axis x line=bottom,
    axis y line=left,
    enlarge x limits=0.12,
  x axis line style={-},
  }
}

\begin{groupplot}[
  group style={
    group size=4 by 1,
    horizontal sep=0cm,
    ylabels at=edge left,
    yticklabels at=edge left,
  },
  width=0.38\linewidth,
  height=0.60\linewidth,
  sidebysidecap,
]

% =================================================================
% Denmark
% =================================================================
\nextgroupplot[
  xlabel={Denmark},
  ylabel style={align=center},
  ylabel={Annualised capital\\expenditure [\unit{\euro\,bn/year}]},
  legend to name=combinedcaplegendtwo,
  legend columns=8,
  legend cell align=left,
  legend style={
 draw=none, fill=none, font=\small, nodes={scale=0.95, transform shape},
    /tikz/every even column/.append style={column sep=0.1em},
  },
]

% CAPEX bars (shifted right, solid)
% Read from capex_by_tech_DK.csv
% Technology order: Nuclear, Offshore Wind, Onshore Wind, Solar, Hydrogen
\pgfplotstableread[col sep=comma]{capex_by_tech_DK.csv}\capexDK
\addplot[fill=cNuclear, draw=none] table[x=Year, y=Nuclear] {\capexDK}; \addlegendentry{Nuclear}
\addplot[fill=cOffwind, draw=none] table[x=Year, y={Offshore Wind}] {\capexDK};     \addlegendentry{Offshore wind}
\addplot[fill=cOnwind, draw=none] table[x=Year, y={Onshore Wind}] {\capexDK}; \addlegendentry{Onshore wind}
\addplot[fill=cSolar, draw=none] table[x=Year, y={Solar}] {\capexDK}; \addlegendentry{Solar}
\addplot[fill=cHydrogen, draw=none] table[x=Year, y={Hydrogen}] {\capexDK};   \addlegendentry{Hydrogen}

% =================================================================
% Finland
% =================================================================
\nextgroupplot[
  xlabel={Finland},
  axis y line=left, y axis line style={opacity=0}, ytick style={draw=none},
]

% CAPEX bars (shifted right, solid)
% Read from capex_by_tech_FI.csv
\pgfplotstableread[col sep=comma]{capex_by_tech_FI.csv}\capexFI
\addplot[fill=cNuclear, draw=none] table[x=Year, y=Nuclear] {\capexFI};
\addplot[fill=cOffwind, draw=none] table[x=Year, y={Offshore Wind}] {\capexFI};
\addplot[fill=cOnwind, draw=none] table[x=Year, y={Onshore Wind}] {\capexFI};
\addplot[fill=cSolar, draw=none] table[x=Year, y={Solar}] {\capexFI};
\addplot[fill=cHydrogen, draw=none] table[x=Year, y={Hydrogen}] {\capexFI};

% =================================================================
% Norway
% =================================================================
\nextgroupplot[
  xlabel={Norway},
  axis y line=left, y axis line style={opacity=0}, ytick style={draw=none},
]

% CAPEX bars (shifted right, solid)
% Read from capex_by_tech_NO.csv
\pgfplotstableread[col sep=comma]{capex_by_tech_NO.csv}\capexNO
\addplot[fill=cNuclear, draw=none] table[x=Year, y=Nuclear] {\capexNO};
\addplot[fill=cOffwind, draw=none] table[x=Year, y={Offshore Wind}] {\capexNO};
\addplot[fill=cOnwind, draw=none] table[x=Year, y={Onshore Wind}] {\capexNO};
\addplot[fill=cSolar, draw=none] table[x=Year, y={Solar}] {\capexNO};
\addplot[fill=cHydrogen, draw=none] table[x=Year, y={Hydrogen}] {\capexNO};

% =================================================================
% Sweden
% =================================================================
\nextgroupplot[
  xlabel={Sweden},
  axis y line=left, y axis line style={opacity=0}, ytick style={draw=none},
]

% CAPEX bars (shifted right, solid)
% Read from capex_by_tech_SE.csv
\pgfplotstableread[col sep=comma]{capex_by_tech_SE.csv}\capexSE
\addplot[fill=cNuclear, draw=none] table[x=Year, y=Nuclear] {\capexSE};
\addplot[fill=cOffwind, draw=none] table[x=Year, y={Offshore Wind}] {\capexSE};
\addplot[fill=cOnwind, draw=none] table[x=Year, y={Onshore Wind}] {\capexSE};
\addplot[fill=cSolar, draw=none] table[x=Year, y={Solar}] {\capexSE};
\addplot[fill=cHydrogen, draw=none] table[x=Year, y={Hydrogen}] {\capexSE};

\end{groupplot}

% --- shared legend below the groupplot ---
\node[anchor=north, yshift=0cm, xshift=0.1cm]
  at (current bounding box.south) {\pgfplotslegendfromname{combinedcaplegendtwo}};

\end{tikzpicture}
  \vspace*{-0.7cm}
\caption{Annualised capital expenditures by technology and year for each Nordic country from 2030.
Denmark sees the largest investment in the Nordics, with a strong focus on offshore wind and solar power, as expected from the NGDP and simulations.
}\label{fig:4}
\vspace*{-0.3cm}
\end{figure}
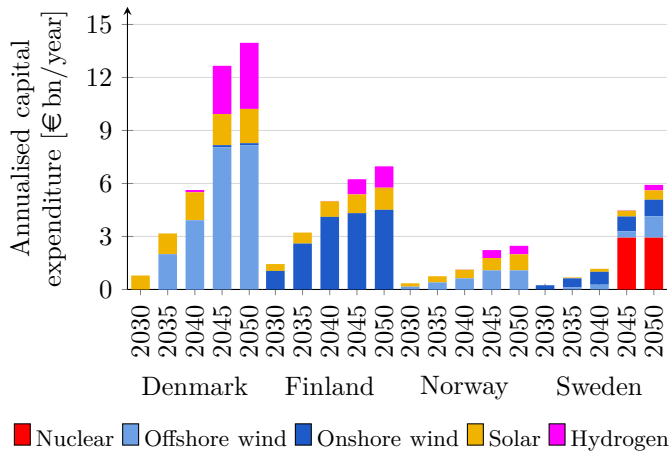

In Fig.~\ref{fig:4}, the annualised capital expenditures for the generation and hydrogen technologies in the PyPSA-Eur optimised scenario are shown, given in billions of euros per year.
Denmark's large expansion of offshore wind and solar power, along with the strong late hydrogen deployment, are the main drivers of the capital expenditure in the Nordic system, raising to up to \qty{8.17}{\euro\,bn/year} for offshore wind power, \qty{1.94}{\euro\,bn/year} for solar, and \qty{3.74}{\euro\,bn/year} for hydrogen in 2050.
The capital expenditures are centrally due to the generation technologies, with hydrogen technologies having smaller costs.
Norway's capital expenditure are minimal, suggesting the system is adequate to meet the demand with the existing generation capacity, with only a small expansion of solar power and hydrogen technologies.
Sweden's potential nuclear expansion will come at a considerable cost in 2045, which casts doubt on whether this is a likely pathway for the country.

We note moreover that the PyPSA-Eur simulations paint a picture wherein Norway moves from exporting roughly \qty{12}{TWh} of electricity in 2025 to importing between \qty{2}{TWh} and \qty{5}{TWh} in 2050.
Denmark moves from being a net importer of roughly \qty{13}{TWh} in 2025 to being a net exporter of up to \qty{37}{TWh} in 2045 and \qty{15}{TWh} in 2050.
The PyPSA-Eur simulation places Finland as a net exporter of \qty{8}{TWh} in 2025, which is a clear contrast with the countr\revise{y'}s importing profile, which reduced considerabl\revise{y} in the past 5 years between 2020 and 2025, but i\revise{t} remains a net importer.
The simulation paints Finland remaining a net exporter all the way to 2050.
We note that this agrees with the mismatch of electricity consumption in Finland evident in Fig.~\ref{fig:2}, and points to an underestimation of the electricity demand in Finland in the PyPSA-Eur model.
\revise{We remark, nevertheless, that the simulated decreasing electricity consumption by the distribution grid is unnatural, both for Finland and Denmark, as it points to a substantial decrease in electrical consumption. PyPSA-Eur explicitly decouples heating (heat pumps and central and decentral heating), but halving the electrical needs in Denmark and Finland between 2040 and 2050 is doubtful.}
Nonetheless, the simulation speaks to an overall exporting profile for the Nordic system, totalling almost \qty{40}{TWh} in 2050, while peaking at \qty{55}{TWh} in 2045.

\section{Conclusions and Outlook}

In this work, we contrast the Nordic Grid Development Perspective (NGDP)~\cite{NGDP2025} with a least-cost optimisation of the Nordics and North Sea-centric European electricity system using the PyPSA-Eur model.
A central disagreement between the predictions by the Nordic TSO emerges with the timing and system role assigned to hydrogen technologies.
In the NGDP, hydrogen appears early and grows steadily from 2030 onward, especially in Denmark, Finland, and Sweden, implying that electrolysis is treated as a near-term pillar of industrial development and grid planning.
In contrast, the least-cost PyPSA-Eur pathway delays substantial hydrogen uptake until 2045--2050, where it emerges mainly as a late-stage flexibility option once variable renewable generation is already large and hydrogen costs have fallen sufficiently.
This is a meaningful strategic contrast: the NGDP reads hydrogen as an enabling infrastructure that must be prepared ahead of demand, whereas PyPSA-Eur reads hydrogen as an option whose value appears only after a strongly VRE-dominated system has been established.
Our results, therefore, suggest that early hydrogen build-out is not a robust outcome of cost optimisation alone, but rather a planning choice that depends on expectations about industrial uptake, coordination, and future market formation.
The expectation is that the existing system can still manage the power-energy balance, likely due to the large existing hydroelectricity and nuclear generation, and that this setup is sufficient throughout 2040.
Hydrogen technologies only become cost-effective as a flexibility and seasonal storage solution in a highly VRE-dominated system from 2045.

This mismatch between the NGDP and PyPSA-Eur model should not be interpreted too narrowly as the model disproving the NGDP, but rather as a reflection of the possible optimism around hydrogen technologies emerging very fastly in the next decade, as needed by the NGDP.
The two approaches embed different assumptions about demand growth, sector definitions, and infrastructure needs, and these differences are particularly important for hydrogen.
PyPSA-Eur does not fully represent several demand categories emphasised in the NGDP, including data centres and parts of transport demand, and appears to underestimate industrial electricity growth in all four countries.
Under such conditions, early hydrogen deployment is less needed.
Conversely, the NGDP may overstate the pace at which hydrogen demand will materialise if announced projects, industrial ambitions, and grid-development expectations do not translate into realised load.
The main conclusion is therefore that hydrogen is the key uncertainty in the Nordic transition, not yet a settled outcome and that between the two predictions, NGDP appears optimistic and PyPSA-Eur appears pessimistic about the near-term role of hydrogen.

Future work should consider, within the limitations of the optimisation model, the inclusion of more detailed demand categories, such as data centres and transport demand, higher temporal and spatial resolution, which inherently yields a more detailed representation of the internal transmission grid, and a clearer representation of the industrial demand aligning with the Nordic governments' targets.
Hydrogen technologies' roles need to be contrasted strongly with their expected costs to ensure their representation is faithful.

\section*{Acknowledgment}
The authors acknowledge the valuable contributions of Mahnoor Saleem, Vetle Malum, and Jonas Jortveit Andresen to the modelling and data analysis carried out in previous models relevant to this work.

\printbibliography

\end{document}